\documentclass[11pt]{article}
\usepackage[colorlinks=true,citecolor=blue,linkcolor=blue]{hyperref}
\usepackage{multirow}
\usepackage[left=2cm,right=2cm,top=2cm,bottom=2cm]{geometry}
\usepackage{cite}
\usepackage{diagbox}
\usepackage{psfrag}
\usepackage{mathtools}
\usepackage{graphicx}
\usepackage{epsfig}
\usepackage{verbatim}
\usepackage{booktabs}
\usepackage{amsmath,amsfonts,amssymb}
\usepackage{pstcol,pst-fill,pst-grad}
\usepackage{pstricks,pst-fill,pst-grad}
\usepackage{euscript}
\usepackage{pstricks}
\usepackage{tabularx}
\usepackage{wrapfig}
\usepackage{slashed}
\usepackage[toc,page]{appendix}
\usepackage{here}
\usepackage{multirow}
\usepackage{ulem}

\date{}
\begin{document}
	\title{\vspace{-3cm}
		\hfill\parbox{4cm}{\normalsize \emph{}}\\
		\vspace{1cm}
		{Enhancement in physical properties  of Pb-Based Perovskite Oxides ($PbGeO_{3}$): Ab initio Calculation
	}}
	\vspace{2cm} 
	
	\author{M. Agouri$^{1}$, A. Waqdim$^{1}$, A. Abbassi$^{1,}$\thanks{Corresponding author, E-mail: abbassi.abder@gmail.com}, M. Ouali$^{1}$, S. Taj$^{1}$, B. Manaut$^{1,}$\thanks{Corresponding author, E-mail: b.manaut@usms.ma}, M. Driouich$^1$ \\
			{\it {\small$^1$ Laboratory of Research in Physics and Engineering Sciences,}}\\
			{\it{\small Sultan Moulay Slimane University, Polydisciplinary Faculty, Beni Mellal, 23000, Morocco.}}\\
		}	
		\maketitle \setcounter{page}{1}
		\date{\today}
		
		\begin{abstract}
In the present paper, the electronic, structural, thermodynamic, and elastic properties of cubic $PbGeO_3$ perovskite oxide are presented and computed using the WIEN2k code. 
The structural properties have been evaluated and they are in good agreement with the theoretical and experimental data. A phonon dispersion is made and it reveals that the cubic $PbGeO_3$ perovskite is dynamically stable. In addition, the electronic properties of $PbGeO_3$ shows an opening gap energy, meaning a semiconductor behavior with an indirect band gap equal to $1.67\;eV$. Moreover, the obtained elastic constants of cubic $PbGeO_3$ satisfy Born’s mechanical stability criteria, and this indicates that our compound behaves as a stable ductile material. The temperature-pressure effects on thermodynamic parameters are investigated using the Gibbs2 package. Finally, based on the obtained results about the cubic $PbGeO_3$ perovskite properties, we assume that this compound will have potential applications.
		\end{abstract}
		
		\maketitle
		\newpage
\section{Introduction}
Since the discovery of $CaTiO_3$ \cite{1}, the perovskite oxides family has been a major subject of interest, and this is mainly due to its multi-functional character \cite{2,3}. It has received great attention for its exploitation in many applications such as solar cells \cite{4}, spintronic and optoelectronic \cite{5,6}. The exploitation of these materials mainly depends on their flexible structure, variable formula, and also their various properties \cite{7,8}. Consequently, a lot of experimental and theoretical studies have proved that the perovskite oxides family has unique physical properties like photoelectric \cite{9}, magnetic \cite{10}, ferroelectric \cite{11},.. etc.\\ 
Pb-based perovskite oxides have long been investigated for their rich and interesting properties. They are considered proper candidates in energy conversion such as in piezoelectric and ferroelectric devices \cite{12}. Recently, several experimental and theoretical studies have been interested on Lead-based perovskite oxides. The unit cell compounds of $PbXO_3\;(X=Ti,\;V)$ were synthesized, and their crystal structures are determined using Neutron and X-ray diffraction \cite{13,14}. In addition, these compounds have largely been used in ferroelectric and optical sensors \cite{15,16}. The $PbZr_{x}Ti_{1-x}O_3\;(x=0,\;0.4,\;0.6,\;1)$ were prepared and investigated, proving the exploitation of these compounds in optical applications \cite{17}. The structural, magnetic, and electronic properties of tetragonal structures ($Pbmm$ and $P4/mmm$) of  $PbMnO_3$ were calculated theoretically using the density functional theory (DFT) approach \cite{18}. Besides, J.B. Goodenough \textit{et al}., showed in \cite{19} the varied roles of $Pb$ in transition-metal $PbTMO_3\;(TM = V,\; Mn,\; Ni,\; Mn,\; Ti,\; Fe,\; Ru)$ perovskites.\\

Due to its excellent and unique physical properties, Cubic $PbGeO_3$ perovskite oxide have received more attention in both experimental and theoretical physics. In addition, the $PbGeO_3$ perovskite crystallizes in cubic structure which is reported in experimental and theoretical works \cite{20}. 
Using X-ray photoelectron spectroscopy, they have calculated the binding energy of the $PbGeO_3$ and $Pb_{5}Ge_{3}O_{11}$ phases and showed their optical transmission characteristics \cite{11}. Other researchers showed that $PbGeO_3$ is considered an interesting choice for lithium batteries \cite{21,22}. Theoretically, optoelectronic and thermoelectric properties of cubic $PbGeO_3$ were evaluated within the DFT approach \cite{23}.\\
In this paper, we investigate the electronic, structural, thermodynamic and elastic properties of cubic $PbGeO_3$ perovskite oxide using Full Potential Linearised Augmented Plane Wave (FP-LAPW) method within the DFT approach. 
Our report is structured as follows: We start with computational procedures, and then we analyze and discuss the obtained results about the studied physical properties of $PbGeO_3$ perovskite. Finally, a conclusion of the main results is given in the last section.\\
\section{Computational details}
In this paper, we investigate the physical properties of cubic $PbGeO_3$ perovskite oxide by using the FP-LAPW method within DFT approach \cite{24} as implemented in the WIEN2k code \cite{25}.
Based on the Perdew-Burke-Ernzerhof approximation (PBE-GGA) \cite{26} and modified Becke-Johnson (mBJ) exchange potential \cite{27}, we have studied the structural and electronic properties of the $PbGeO_3$ perovskite oxide. 
The elastic parameters have been determined using ElaStic-1.1 package \cite{28}. The separation energy between core and valence electrons is $-10.0\; Ry$. The number of plane waves is limited by $R_{MT} \times K_{max} = 8$. The $l_{max}$ parameter is taken to be $10$ and the Fourier expanded change density is $G_{max} = 12$. The integration of first Brillouin zone is performed with ($6 \times 6 \times 6$) k-points grid in reciprocal space. The crystal structure is designed using VESTA program \cite{29}.\\
The thermodynamic parameters of the cubic $PbGeO_3$ are determined by using the quasi-harmonic Debye model \cite{30,31}. 
The non-equilibrium Gibbs function $G^{*}(P,V,T)$ is defined by the following equation :
\begin{equation}
G^{*}(P,V,T)= E(V)+PV+ H_{vibration}[\Theta_{D}(V),T],
\end{equation}
where $PV$ represents the constant hydrostatic pressure condition and $E(V)$ is the equilibrium energy per unit cell.\\ 
The $H_{vibration}[\Theta_{D}(V),T]$ denotes the vibrational term, which can be written as:
\begin{equation}
	H_{vibration}\left(\Theta_{D}(V),T\right)=mK_{B}T \left\lbrace \dfrac{9\Theta_{D}}{8T}+3 \ln \left(1-e^{-\Theta_{D}/T}\right)-D\left(\frac{\Theta_{D}}{T}\right)\right\rbrace,
\end{equation} 
where $m$ represents the number of atoms per formula and $D\left(\frac{\Theta_{D}}{T}\right)$ is the Debye integral.\\
The Debye temperature $\Theta_{D}$ is expressed as :
\begin{equation}
\Theta_{D}=\dfrac{\hbar}{K_{B}}\left(6\pi^{2}V^{1/2}m\right) ^{1/3} f(\sigma) \sqrt{\dfrac{B_{s}}{M}},
\end{equation}
where $M$ stands for the molecular mass per unit cell. \\
The adiabatic bulk modulus $B_{T}$ is approximately defined as the static compressibility :
\begin{equation}
B_{T} \approx B(V)=V\dfrac{d^{2}E(V)}{dV^{2}}.
\end{equation}
The $G^{*}(P,V,T)$ function is minimized with respect to the volume $V$ as:
\begin{equation}
 \left[\dfrac{dG^{*}(P,V,T)}{dV}\right]_{P,T}.
\end{equation}
By solving the equation (5), we find the thermal equation of states $V(P,T)$.\\
The thermal expansion $\alpha$, bulk modulus $B$ and the heat capacity (at volume constant) $C_V$ are successively given by the following equations\cite{30}:
\begin{equation}
	\alpha=\dfrac{\gamma C_{V}}{B_{T}V},
\end{equation}
\begin{equation}
	B_{T}=V\left(\dfrac{d^{2}G^{*}(P,V,T)}{d^{2}V}\right)_{P,T},
\end{equation}
\begin{equation}
	C_{V}=3mk \left[4D\left(\frac{\Theta}{T}\right) - \dfrac{3\Theta/T}{e^{\Theta/T}-1}\right].
\end{equation}
\section{Results and discussions}
\subsection{Structure and stability}
\begin{figure}[H]
	\centering
	\includegraphics[scale=0.07]{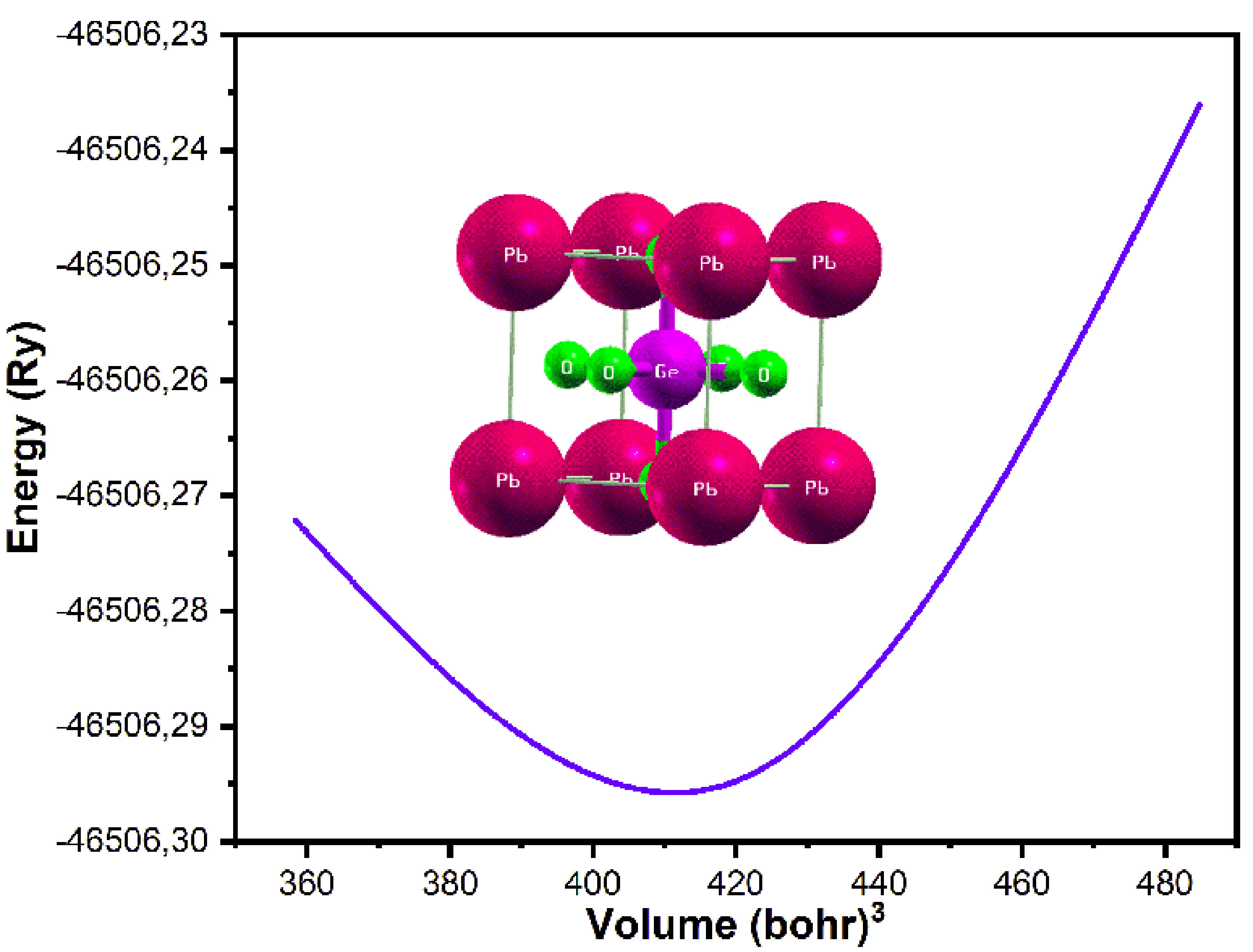}
	\caption{The crystal structure and the optimization plot of $PbGeO_3$.}
	\label{Figure: 1}
\end{figure}

\begin{figure}[H]
	\centering
	\includegraphics[scale=0.15]{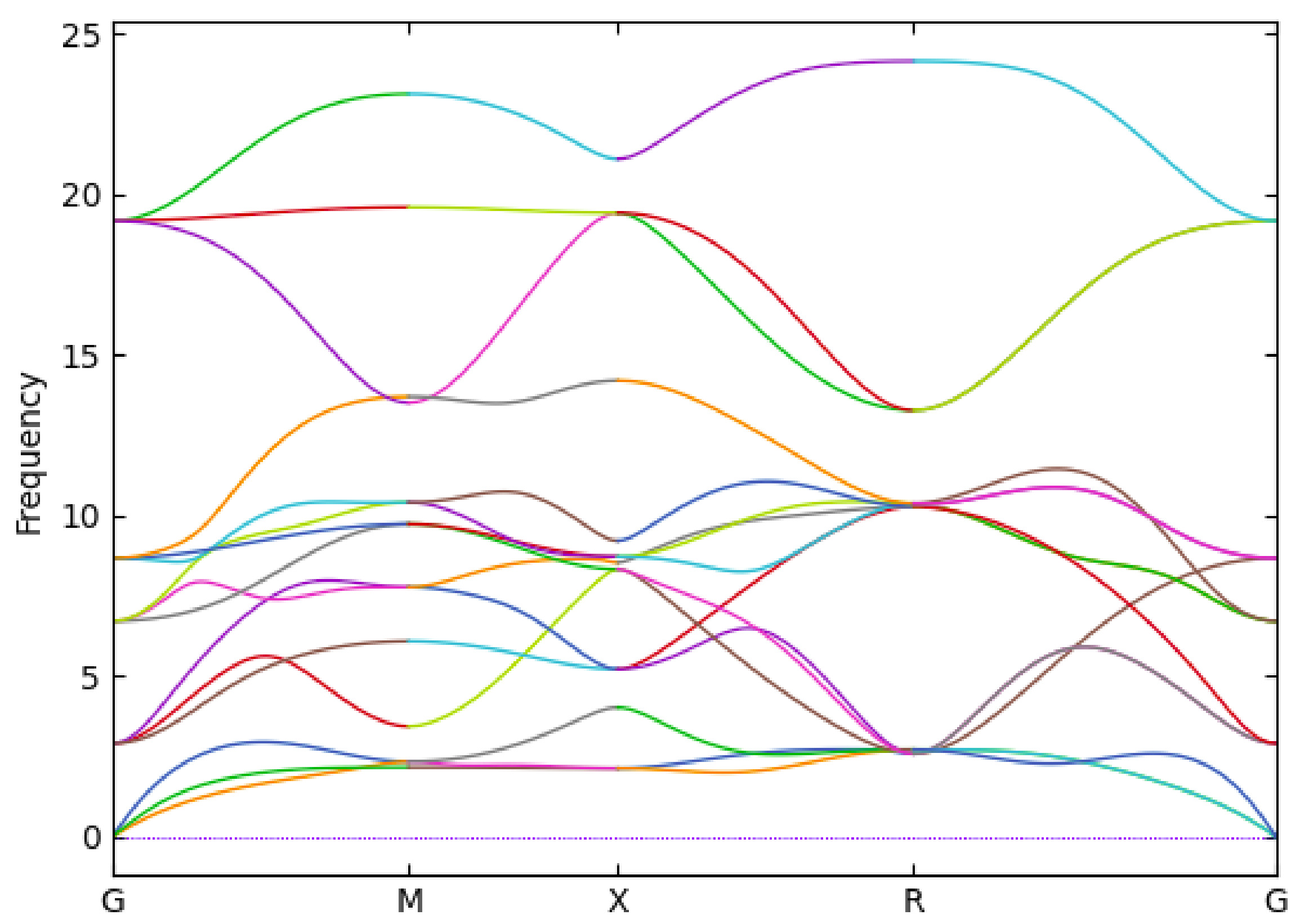}
	\caption{Phonon dispersion curve of $PbGeO_3$.}
	\label{Figure: 2}
\end{figure}

Perovskite oxide $PbGeO_3$ was first optimized based on the experimental lattice parameter. $PbGeO_3$ has an ideal cubic phase with a space group $Pm3m$. The atomic coordinates of the primitive cell of cubic $PbGeO_3$ are defined as $Pb:\;(0,0,0)$, $Ge:\;(1/2,1/2,1/2)$ and $O:\;(0,1/2,1/2)$. Figure \ref{Figure: 1} shows the variation of the total energy as a function of the unit cell volume in addition to the flexible structure of cubic $PbGeO_3$.
The calculated values of lattice constant ($a$) and bulk modulus ($B$) of our compound are summarized in table \ref{Table: 1}. We notice that the obtained results are in good agreement with the theoretical and experimental works \cite{20,23}.
\begin{table}[H]
\centering
\begin{tabular}{ p{2cm} p{5cm} p{5cm} p{3cm}}
\hline 
Compound    & $a$(\AA)            & $B(GPa)$          & Methods \\
\hline
$PbGeO_3$   & 3.8984              & 152.5279          & Our work       \\
			& 3.9680      &                   & Exp  \cite{20}          \\
			& 3.8320, 3.8420, 3.9002,  &198, 157.8647, 180.7519, &Theory\cite{20,23}\\
			&3.8404, 3.8536, 3.8150.   &181.9745, 201.1913.              & \\
\hline
\end{tabular}
\caption {Calculated lattice constant ($a$) and Bulk modulus ($B$) of $PbGeO_3$ compound.}
\label{Table: 1}
\end{table}
In order to examine the dynamic stability of cubic $PbGeO_3$, we have calculated the phonon dispersion using the supercell method within phonopy code \cite{32}. 
Figure \ref{Figure: 2} presents the phonon dispersion of our material $PbGeO_3$. According to this figure, the phonon dispersion curve of our compound shows positive frequencies along the high symmetry directions, indicating the dynamic stability of the cubic $PbGeO_3$ phase.
\section{Electronic properties}
\begin{figure}[H]
	\centering
	\includegraphics[scale=0.1]{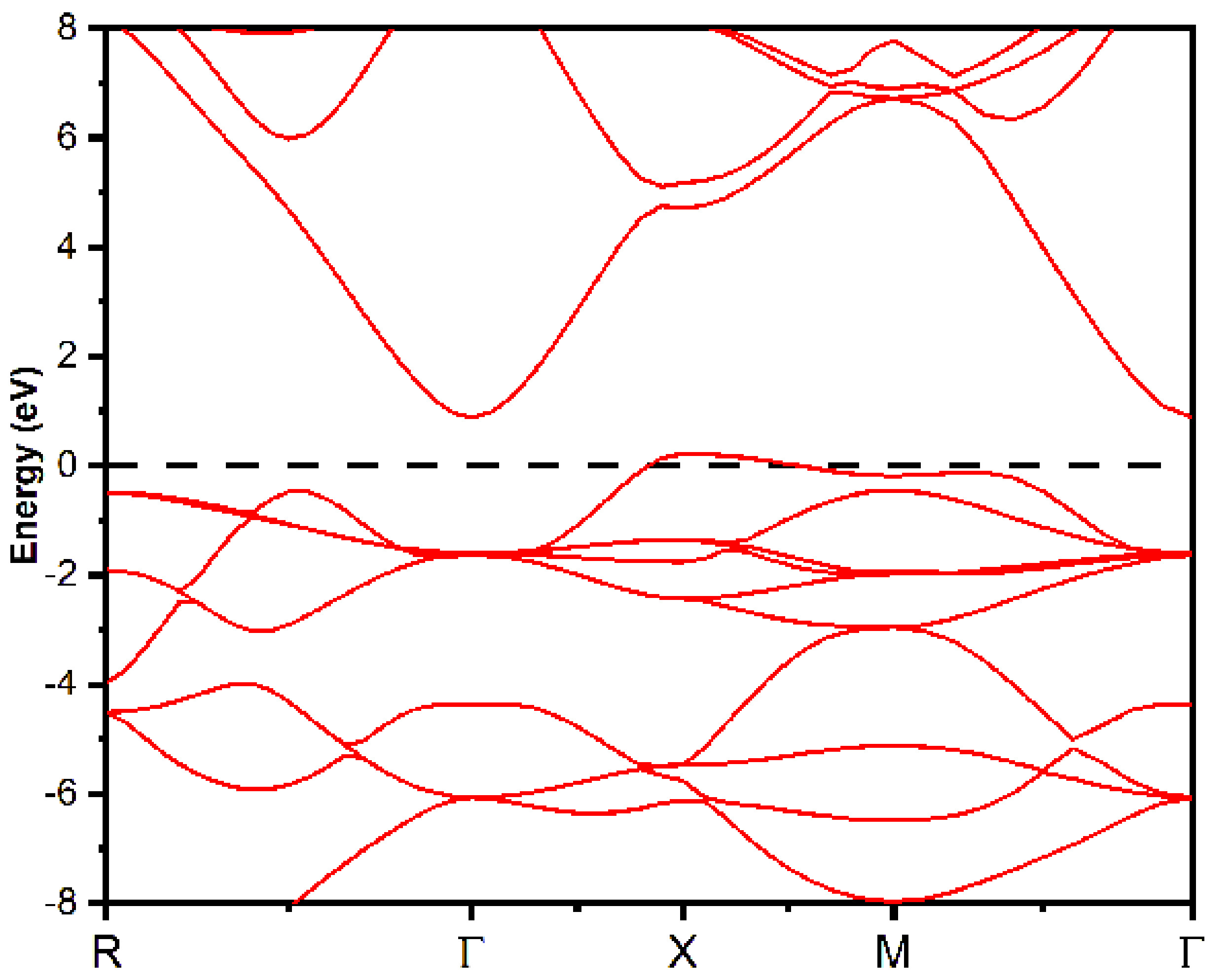}
	\caption{Band structure of $PbGeO_3$ using mBJ approach.} 
	\label{Figure: 3}
\end{figure}

\begin{figure}[H]
	\centering
	\includegraphics[scale=0.1]{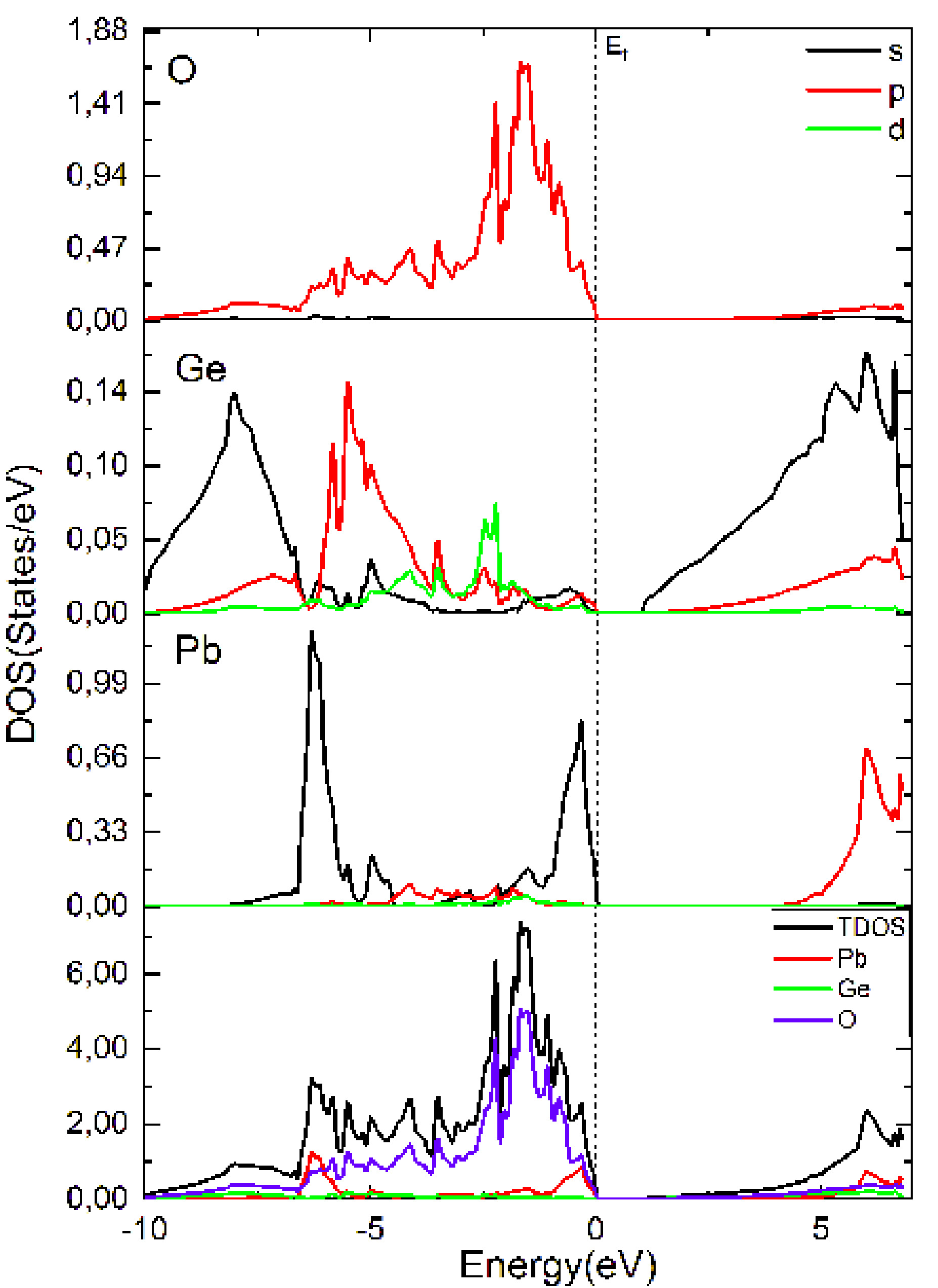}
	\caption{The partial and total density of the states of $PbGeO_3$ using the mBJ approach.} 
	\label{Figure: 4}
\end{figure}

Regarding the electronic properties of cubic $PbGeO_3$, we have investigated the band structure, partial and total density of state using PBE $+$ TB-mBJ exchange-correlation potential. 
We mention that, the mBJ approach gives a large band gap energy, and it solves the problem of underestimation band gap energy \cite{27}. \\
Figure \ref{Figure: 3} shows the obtained electronic band structure of $PbGeO_3$ along the high symmetry directions using the mBJ approach. We can see from this figure that $PbGeO_3$ shows a semiconductor behavior. In addition, we note that the valence band maximum (VBM) and the conduction band minimum (CBM) are placed, respectively, at $X$ and $\Gamma$ points. This means that $PbGeO_3$ has an indirect band gap equal to $1.67\;eV$ ($\Gamma$-X). Moreover, our result is consistent with other theoretical calculation (A.Day \textit{et al.} \cite{23}).
For a better illustration of the contribution of different band energies in the band structure, we have calculated the partial and total density of states as presented in figure \ref{Figure: 4}. We note that the lower region of the valence band consists of all orbitals such as $s-Pb$, $sp-Ge$ and $p$ of Oxygen with hybridization between $Pb$ and Oxygen in TDOS. Near the Fermi level, the $p-$Oxygen is mixed with $s-Pb$ which represents a strong hybridization of $p$ and $s$ of Oxygen and Lead, respectively. 
The gap energy is clearly shown due to the contribution of the $s-Ge$ and $p$ orbitals of Oxygen. For the conduction band (CB), a large dispersion of $s$ orbital of $Ge$ is observed, and it forms a covalent chemical bond with the p orbital of oxygen. 
Similar results are shown in the case of $SrGeO_3$ \cite{33}. Besides, we observe a hybrid state between the $p$ orbitals of $Pb$ and Oxygen around $5\;eV$.
\subsection{Elastic properties}
Elastic constants play a significant role in material engineering as it provides important information about the ductility, brittleness, stillness, and also the mechanical stability of materials \cite{34}. The elastic properties of cubic $PbGeO_3$ are investigated using the ElaStic1-1 package \cite{28}.
$PbGeO_3$ is a cubic crystalline structure, which has three independent elastic constants ($C_{11}$, $C_{12}$ and $C_{44}$). Table \ref{Table:2} shows the calculated elastic constants of our compound using the modified Becke Johnson (mBJ) approximation. We note that $C_{11}$ represents the longitudinal distortion (compression) and describes the hardness of the material, $C_{12}$ is based on the transverse distortion (compression) and $C_{44}$ represents the resistance to shear deformation, which is based on the shear modulus \cite{34}. \\
The mechanical stability is verified using Born's criteria that are given by \cite{35}:
\begin{equation}
	\begin{aligned}  
		\begin{dcases*}
			C_{11} - C_{12} >0, \\
			{C_{44} >0}, \\
			C_{11} + 2C_{12} >0. 
		\end{dcases*}
	\end{aligned}
\end{equation}
From Table \ref{Table:2}, the calculated elastic constants respect Born's criteria, and this clearly means that cubic $PbGeO_3$ is mechanically stable. From table \ref{Table:2}, one can see that the longitudinal distortion value is higher than $C_{44}$ value, meaning a weak resistance to the shear modulus compared to the longitudinal distortion \cite{36}.
\begin{table}[H]
\centering
\begin{tabular}{ |p{3cm}|p{3cm}|p{3cm}|p{3cm}|  }
	\hline
	\multicolumn{4}{|c|}{\textbf{Elastic Coefficients}}  \\
	\hline
	Coefficient     & C$_{11}$   &  C$_{12}$  & C$_{44}$\\
	\hline
	PbGeO$_3$        & 153.4 & 141.1 & 115.5\\
	\hline
\end{tabular}
\caption {The obtained elastic constants of the $PbGeO_3$ compound.}
\label{Table:2}
\end{table}
Based on elastic constant $C_{\alpha\beta}$, The mechanical constants such as shear modulus ($G$), bulk modulus ($B$), Cauchy pressure ($C"$), Pugh's ratio ($B/G$), Poisson's ratio ($\nu$), anisotropy ($A$) are calculated through the VRH (Voigt-Reuss-Hill) approximation by using the following formulas \cite{37,38}:
\begin{equation}
	B=\dfrac{2C_{12}+C_{11}}{3}, \hspace*{0.5cm} G=\dfrac{G_{R}+G_{V}}{2}, \hspace*{0.5cm}
	Y=\dfrac{9BG}{3B+G},\hspace*{0.5cm} \nu=\dfrac{3B-2G}{2(G+3B)},
\end{equation}
where $G_{R}$ and $G_{V}$ are successively the shear modulus of Reuss and Voight approaches such that:
\begin{equation}
	G_{V}=\dfrac{C_{11}+3C_{44}-C_{12}}{5},\; \; \hspace*{0.5cm} 
	G_{R}=\dfrac{5C_{44}(C_{11}-C_{12})}{3(C_{11}+C_{12}+4C_{44})}.
\end{equation}
The obtained results of mechanical parameters are listed in table \ref{Table:3}. The bulk and shear modulus can be used to measure the rigidity of materials \cite{37}. The calculated $G$ and $B$ values are $43.03\;GPa$ and $145.18\;GPa$, respectively. 
We notice that the $B$ value found from elastic constants is nearer to that obtained by fitting the Birch-Murnaghan equation of state. This comparison ensures that our computed elastic constants are correct. In addition, Young's modulus $Y$ also deals with the hardness or stillness of materials \cite{39}. The obtained value of $Y$ is equal to $117.49\;GPa$, which is important as it is greater than $100$. Therefore, we assume that $PbGeO_3$ behaves as a hard material. Anisotropy ($A$) is also an important parameter in industrial science for detecting micro-cracks in materials \cite{40}. From table \ref{Table:3} the calculated value of $A$ is $0.6$, meaning that $PbGeO_3$ shows an anisotropy aspect.\\
The Poisson's ratio ($\nu$), Cauchy pressure ($C"$) and Pugh's ratio ($B/G$) reveal the ductile or brittle aspect of materials. It is well known that the critical value of Pugh's ratio ($B/G$) which separates the ductile/brittle aspect is found to be $1.75$.
Consequently, a material will show a ductile aspect for values of $B/G$ higher than $1.75$, while it shows a brittle nature for values of $B/G$ less than this critical value \cite{41}. 
Another index of brittleness/ductility is the Cauchy pressure ($C"$). 
The positive and negative values of $C"$ are, respectively, related to the ductility and brittleness nature \cite{42}. The parameter of Poisson's ratio ($\nu$) is also a significant factor to distinguish the ductility/brittleness of materials with its critical value which is 0.26. The material will be ductile (brittle) when the Poisson's ratio is higher (less) than 0.26 \cite{43}. Based on these roles and the obtained results, we conclude that $PbGeO_3$ shows a ductility aspect.
\begin{table}[H]
	\centering
\begin{tabular}{ |p{1.5cm} p{1.5cm} p{1.5cm} p{1.5cm} p{1.5cm} p{1.5cm} p{1.5cm} p{1.5cm}|  }
	\hline
	\multicolumn{8}{|c|}{\textbf{Mechanical Constants}}  \\
	\hline
	Constant   &$B$(GPa)    &$G$(GPa)   &$B/G$ &$Y$(GPa)     & C$^{"}$(GPa) & $\nu$ & $A$\\
	\hline
	PbGeO$_3$  &145.18 &43.03 &3.37  &117.49     &25.6 & 0.37 & 0.67\\
	\hline
\end{tabular}
\caption {The obtained values of shear modulus $G$, bulk modulus $B$, Pugh’s ratio $B/G$, Young’s modulus $Y$, Poisson’s ratio $\nu$, Cauchy pressure $C^{''}$ and elastic anisotropic factor $A$ of $PbGeO_3$ compound.}
\label{Table:3}
\end{table}
\subsection{Thermodynamic properties}
\begin{figure}[H]
	\centering
	\includegraphics[scale=0.1]{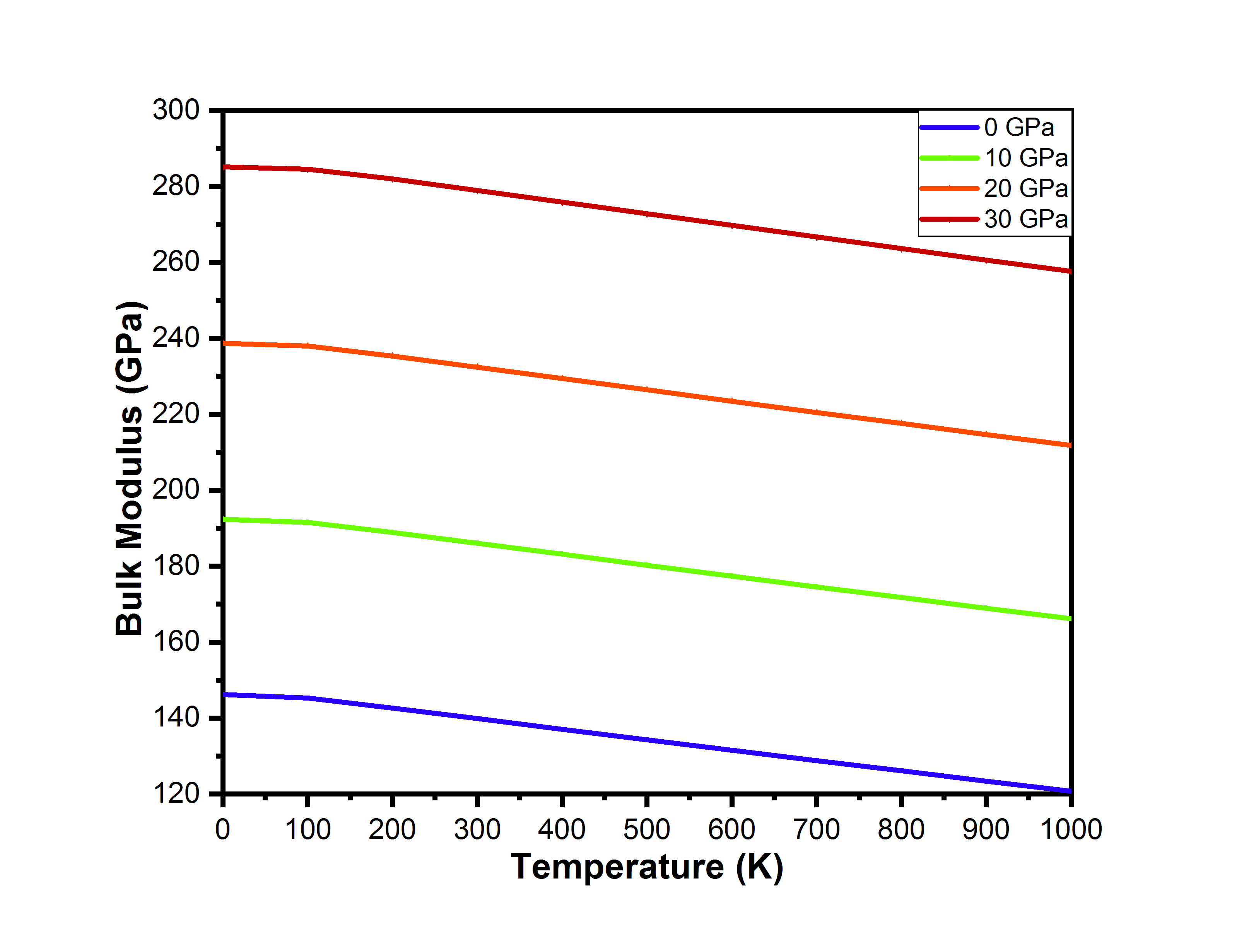}
    \caption{Bulk modulus versus temperature of cubic $PbGeO_3$ perovskite oxide at different pressures.}
   \label{Figure:5}
\end{figure}

The study of thermodynamic properties depending on temperature and pressure provides detailed information about material applications and offers a point of view on their fabrication \cite{44}. \\
The thermodynamic parameters such as bulk modulus ($B$), volume ($V$), thermal expansion ($\alpha$), Deybe temperature ($\theta_{D}$) and heat capacity ($C_{V}$) have been calculated using the Gibbs2 package within the quasi-harmonic approach \cite{30}. The temperature-pressure effects on thermodynamic parameters of cubic $PbGeO_3$ perovskite oxide have been studied in the range of $0$ to $1000\;K$ for temperature and $0$ to $30\;GPa$ for pressure.
\begin{figure}[H]
	\centering
	\includegraphics[scale=0.1]{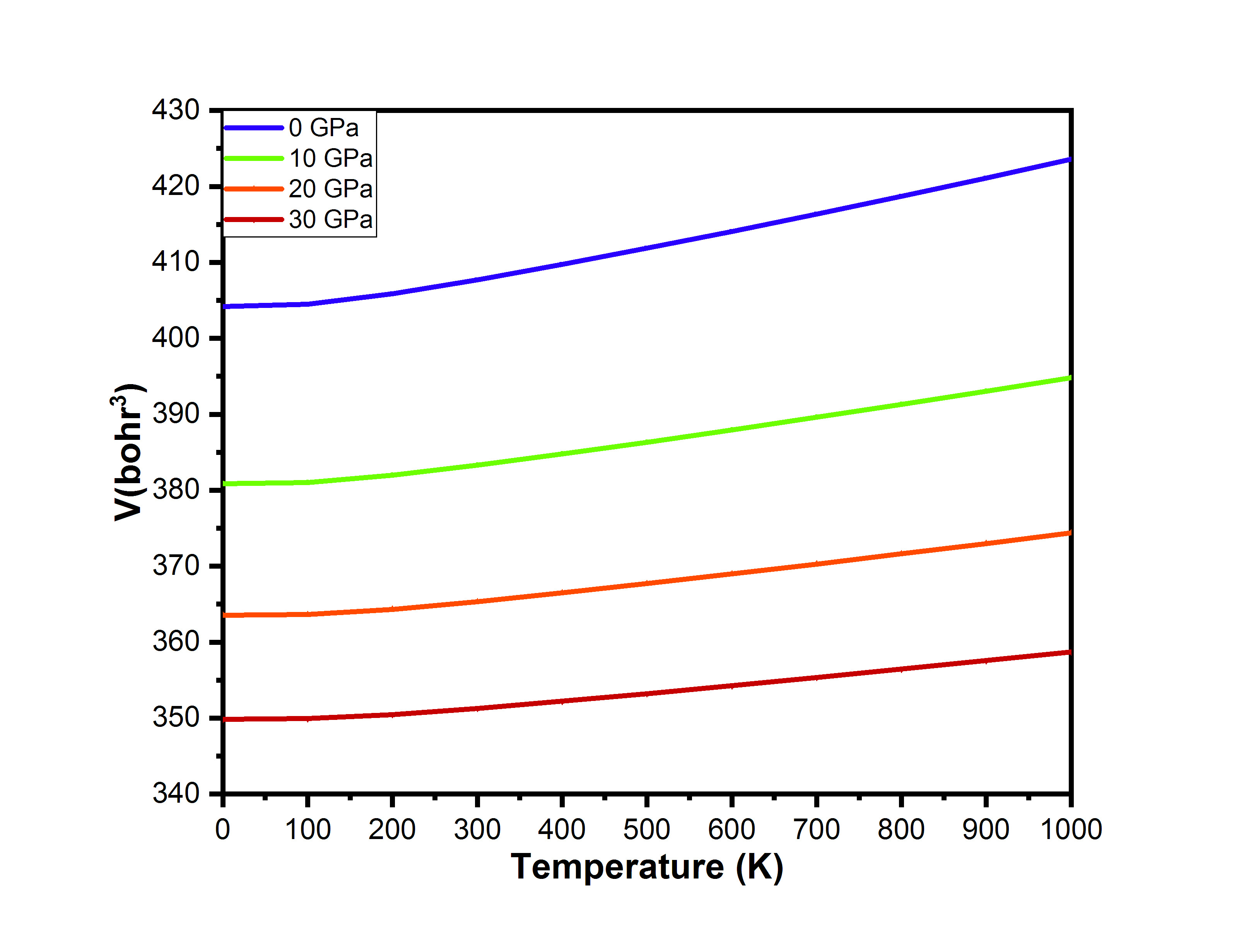}
	\caption{Unit cell volume versus temperature of cubic $PbGeO_3$ perovskite oxide at different pressures.}
	\label{Figure:4}
\end{figure}
The effect of both pressure and temperature on the unit cell volume of cubic $PbGeO_3$ perovskite oxide is illustrated in figure \ref{Figure:4}. We notice that the volume increases with increasing temperature (expansion), and decreases with increasing pressure (compression). We also remark that the calculated value of volume at $0\;GPa$ and $0\;K$ is in good agreement with what we found in the structural data.\\
The variation of bulk modulus $B$ of $PbGeO_3$ versus temperature at certain pressures is shown in figure \ref{Figure:5}.
We notice that the effect of temperature and pressure on the bulk modulus $B$ is opposite to their effect on the volume curve.
 Indeed, the bulk modulus $B$ increases with increasing pressure, and it decreases when we increase the temperature. Therefore, the compressibility decreases with pressure at a certain temperature, while it increases with increasing temperature at a particular pressure \cite{45}.
\subsubsection{Thermal expansion}
\begin{figure}[H]
	\centering
	\includegraphics[scale=0.1]{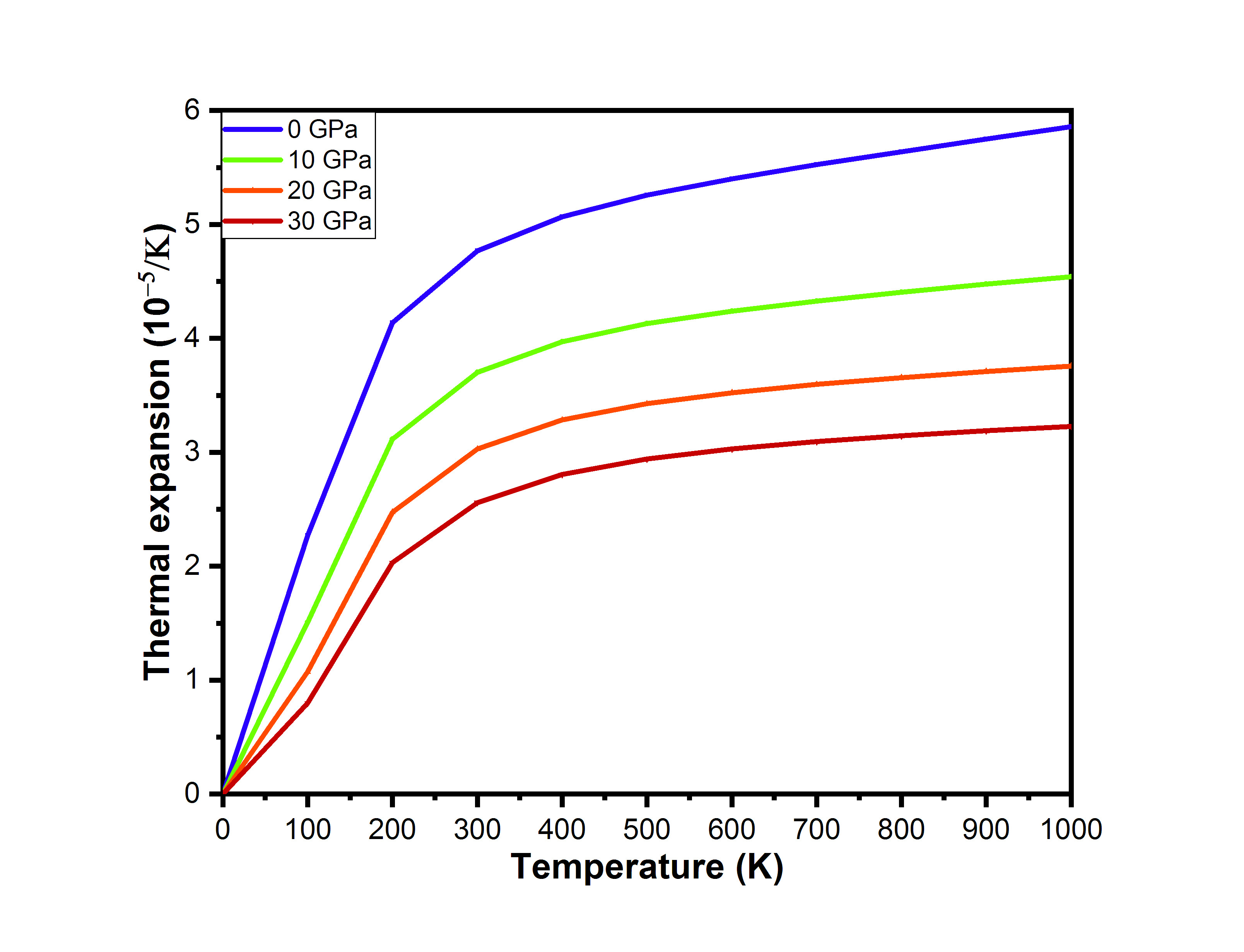}
	\caption{Thermal expansion ($\alpha$) as a function of temperature of cubic $PbGeO_3$ perovskite oxide at different pressures.}
	\label{Figure:6}
\end{figure}

The thermal expansion coefficient ($\alpha$) of cubic $PbGeO_3$ is calculated with respect to temperature at certain pressures as displayed in figure \ref{Figure:6}. It is a significant parameter that provides information about the inter-atomic forces of materials, and it is also linked to the anharmonicity of the lattice interaction potential \cite{46}.\\
From figure \ref{Figure:6}, it can be seen that the expansion coefficient $\alpha$ increases rapidly up to $300\;K$, then we observe a nearly linear increase at higher temperatures. 
On the other side, increasing pressure at a particular temperature leads to a decrease in the thermal expansion coefficient. This result indicates that the $PbGeO_3$ exhibits excellent volume invariance under high pressure \cite{47}. Consequently, increasing pressure and temperature have opposite effects on thermal expansion.
\subsubsection{Debye temperatures}
\begin{figure}[H]
	\centering
	\includegraphics[scale=0.4]{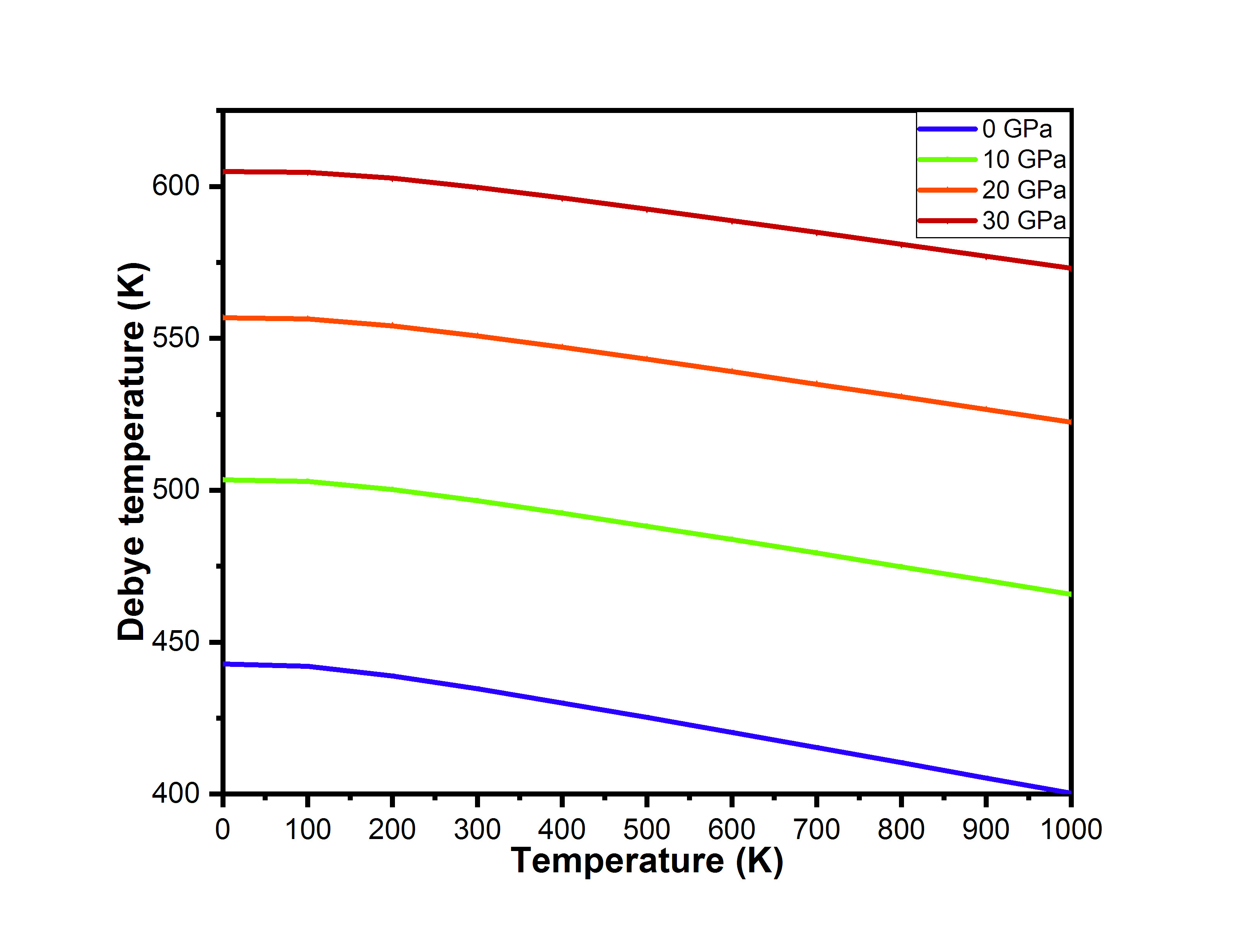}
	\caption{Debye temperature ($\theta_{D}$) as a function of temperature of cubic $PbGeO_3$ perovskite oxide at different pressures.}
	\label{Figure:7}
\end{figure}

The Debye temperature is a special temperature of solids which exhibits the temperature at which the atomic vibrations of material reach their maximum of possible modes \cite{48}. It is a proper estimation of the rigidity of solids \cite{49}.\\
Figure \ref{Figure:7} illustrates the Debye temperature versus temperature for $PbGeO_3$ at given pressures. 
It is obvious that the $\theta_{D}$ curve is approximately constant in the range of $0$ to $100\;K$ for all considered values of pressure.
This result indicates that the crystal experiences a weak anharmonicity and a slight expansion in this temperature range. 
Beyond $120\;K$, the Debye temperature $\theta_{D}$ is reduced gradually with increasing temperature, and this indicates a variation in the atomic vibration spectrum \cite{50}.
\subsubsection{Heat capacity}
\begin{figure}[H]
	\centering
	\includegraphics[scale=0.4]{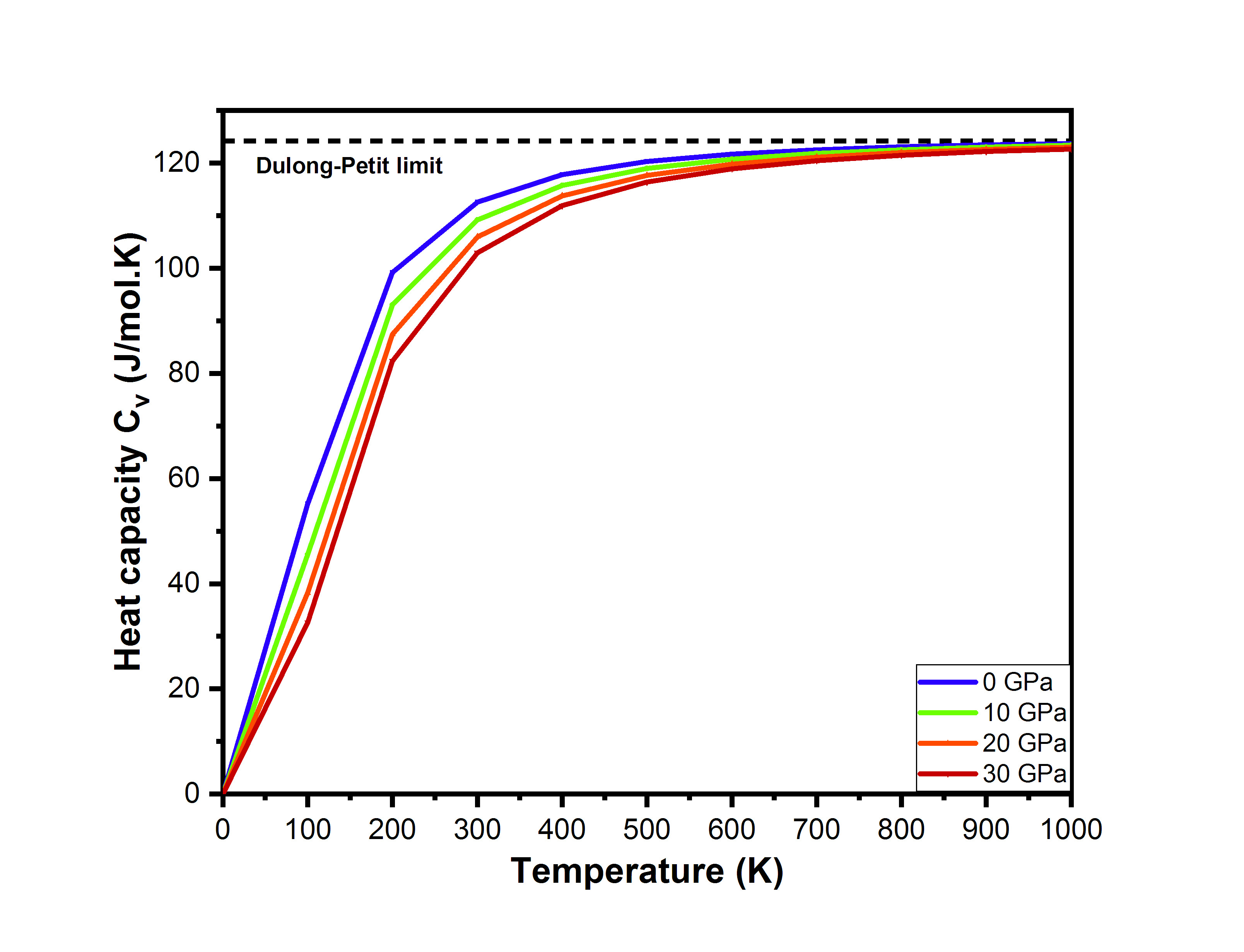}
	\caption{Heat capacity ($C_V$) with respect to temperature of cubic $PbGeO_3$ perovskite oxide at different pressures.}
	\label{Figure:8}
\end{figure}

The heat capacity is not only a significant feature that provides the necessary information about the vibrational characteristics of the lattice but also an obligatory parameter for many applications \cite{51}. \\
The variation of the heat capacity $C_V$ of our compound as a function of temperature is shown in figure \ref{Figure:8}. We notice that the heat capacity presents a similar behavior with different pressures. For a particular pressure, it can be seen that  $C_V$ obeys $T^3$ at low temperatures, which accords with the simple Debye model \cite{52}. 
Then, the $C_V$ increases slowly with increasing temperature, and it approaches the classical Dulong-Petit limit ($3R \times 5$ atoms $= 15R=124.71\; J/K.mol$) at high temperatures. \cite{52}.
\section{Conclusion}
In this paper, we have studied the electronic, elastic, structural, and thermodynamic properties of cubic $PbGeO_3$ based on the FP-LAPW method.  The mechanical and elastic parameters are evaluated, and they prove that the cubic $PbGeO_3$ is mechanically stable. 
In addition, the phonon dispersion shows positive frequencies, confirming that the $PbGeO_3$ compound is dynamically stable. 
Next, we analyze the electronic properties which reveal that cubic $PbGeO_3$ is a p-type semiconductor with an indirect band gap. 
Finally, the thermodynamic parameters such as volume, Debye temperature, bulk modulus, and heat capacity are predicted for the cubic $PbGeO_3$ perovskite using quasi-harmonic Debye approximation with pressure and temperature in the range of $0$ to $25\;GPa$ and $0$ to $1000\;K$, respectively. Overall, the discussed parameters exhibit the good efficiency of our material.

\end{document}